# Tensile testing of cylindrical multi-shell Cu nanowire

Won Woo Kim, Jeong Won Kang*, Tae Won Kim, and Ho Jung Hwang

Semiconductor Process and Device Laboratory, Department of Electronic Engineering, Chung-Ang University, 221 HukSuk-Dong, DongJak-Ku, Seoul 156-756, Korea

Gyu Young Lee

Department of Information and Communication Engineering, Samchok National University, 253 Gyodong, Samchok, Kangwondo 245-080, Korea

We have simulated the tensile testing of cylindrical multi-shell Cu nanowire. Elongated cylindrical multi-shell Cu nanowire was transformed into a pentagonal structure. Pentagonal nanowires are composed of a central atomic strand and pentagonal tubes made of five-times-folded <110>{100} sheets.

*E-mail: kok@semilab3.ee.cau.ac.kr
Tel: 82-2-820-5296
Fax: 82-2-812-5318

# Introduction

The studies on structures, mechanical, and electronic properties of molecular electronic devices, metallic nanowires (NWs), and carbon nanotubes (CNTs) have been intensively performed over the past decade. Until quite recently, ultra-thin NWs of Au [1,2,3], Cu [3, 4], Pb and Al [2, 5], Pt and Ag [6], and Ti [1] have been investigated by using molecular dynamics (MD) simulations. These studies showed that helical, multi-shelled, and filled structures exist for ultra-thin NWs of several fcc metals. Yanson et al. studied multi-shell structured NWs of Na [7]. The stability of Na NWs was studied by modeling them as infinite uniform jellium cylinders, and solving self-consistently [8]. Multi-shell NWs have also been found for several inorganic layered materials, such as $WS_2$, $MoS_2$, and $NiCl_2$ [9,10]. In recent works, long metallic NWs with well-defined structures, several nanometers in diameters, have been fabricated using different methods [11,12,13]. Novel helical multi-shell structures have also been observed in ultrathin Au NWs [11,12].

The various deformations of Cu NWs [15], the stretching of Cu NWs [16] and Au NW [17,18], the strain rate effect induced by amorphization of pure Ni and NiCu alloy NWs [19], and the stretching of several fcc and bcc metals [20] have also been investigated by using atomistic simulations. Reference [20] briefly reviewed the literatures relevant to the MD simulations of tensile testing. In most previous studies of tensile testing, NWs with only single crystalline structure have been initially considered.

However, as mentioned above, it has been found that ultra-thin metallic NWs have the cylindrical multi-shell (CMS) structures. The compression study on the CMS-type Au NW was performed [21]. This investigation performs the tensile testing of a CMS-type Cu NW using atomistic simulations. The structural transformations, tensile force variations, and deformation behavior of CMS-type Cu NW will be discussed.

## Simulation Procedure

In order to describe a CMS-type NW, we used the notation n-n'-n''-n''', when the NW consists of coaxial tubes with n, n', n'', and n''' helical atom rows [11]. In this work, we used the 16-11-6-1 CMS-type NW, which has the diameter of 14 Å and the length of 66 Å. Initial atomic arrangement in the wire was relaxed by the steepest descent (SD) method. The initial structure for the MD simulations was obtained from MD step of 200 000 with timestep 0.5 fs at 300 K. The bottom and the top layer were elongated with $5\times10^{-10}$ Å in the MD simulation and with $5\times10^{-3}$ Å in the SD simulation, respectively, and were fixed during the simulations of each elongation step. The MD methods in our previous works [15,22] were used and the MD timestep was $10^{-7}$ fs. The MD or the SD method calculated the stable atomic configurations after deformations. In the SD simulation, after a small deformation, the stable atomic configuration of NW was relaxed by the SD method. Such small deformation and relaxation were repeated. The tensile forces, which correspond to the force exerted by the pulling agent on the wire, were the attractive force between the top-layer region and the wire region and were calculated after relaxation. In MD simulation, each MD step was applied to each elongation step. Therefore, the elongation velocity was 0.1 m/s, and this is very high stretch velocity compared with STM experiments, $10^{-10}$ m/s. It takes a very long time to simulate stretch velocities of STM experiments by using present computers. Sutton [23] has questioned this dramatic difference between experiment and MD simulation, and Mehrez and Ciraci [16] explained that in spite of this dramatic difference, it appeared that simulations carried out with relatively higher speed could reveal the main features of the atomic rearrangements. If the dramatic differences of stretching velocities between MD simulations and STM experiments are considered, the result using the SD method can be considered as corresponding to cases of very low stretch velocity at low temperature. However, the thermal effects, temperature-dependent deformation mechanisms, cannot be included in the SD simulation.

We have used a many-body potential function of the second-moment approximation of tight-binding scheme [24] that has already been tested in nanocluster and nanowires, etc.

## Results and discussion

Figure 1 shows the tensile force variations as a function of stretching. The tensile force variations show a gradual increase (in quasi-elastic region) to the maximum tensile strength and a rapid drop (in atomistic rearrangement and force relaxation) followed by fracture. The results of the tensile force variation are in a good agreement with the previous results [15,16,17,18,19,20]. At the first yielding, slip of atomic layers was not appeared, but some circular atomic rings in the outer shell of the central region of wire were transformed into polygonal atomic rings. In the other yielding, at least, the perpendicular plane slips and the diagonal direction slips to the wire axis were alternatively appeared in the neck of the wire, and these slip mechanisms made a increase of atomic layers in the wire. The slip caused by an external force preferentially occurs when the least amount of energy is consumed. In bulk fcc materials, this condition is fulfilled when the slips occur in <110> directions on the {111} plane and (1/2)<110>{111} dislocations split into partial dislocations with Burger vectors (1/6)<211> [25]. However, from the previous works using NWs with fcc structure [15,16,17,18], it has been known that the deformation mechanisms of NWs include various dislocation slips, and at least two distinct slip mechanisms mainly affect nanowire deformation processes. One is a glide of a dislocation on {111} planes and the other is a homogeneous slip of one plane of atoms over another plane of atoms. The crossover of the two slip mechanisms is related to reduction of diameter of the nanowire center. Since the CMS-type Cu NWs are composed of {111}-like surface and {111}-like cross-section, our simulations show that the slip mechanisms of the CMS-type NW are similar to that of the previous works using NWs with fcc structure. However, in the structural aspect, whenever yielding occurred, {111}-like surfaces were changed to {100}-like surfaces and atomic layers with the pentagonal cross-section increased in both the MD and the SD simulations. The previous stretching simulations using {100} NWs showed that the {100} NWs have been transformed into structures with {111}-like surface [15]. Since the pentagonal structure is not {100} structure, the result of this paper is not reverse to the results in the previous works. The bottom of Fig. 1 shows atomic structures when the wire is stretched up to 7 Å, and the squares indicate the region with a pentagonal structure with {100}-like surfaces. The upper right side of Fig. 1 shows a unit cell of a pentagonal structure along the wire axis, and the spheres of same color indicate atoms in the same layer. This is a fivefolding multi-shell-type in decahedron model. In the region transformed into the pentagonal structure, displacements due to shear stresses were shown but yieldings and structural rearrangements were never occurred. Since the layers of both ends have been fixed during simulations, a central part of the NW transformed into stable fivefold decahedral packing. Even after breaking of the wire, the pentagonal structures of the wire were maintained. Lisiecki *et al*. [13] have shown that the structures of composed Cu nanorods can be explained by decahedron model. Hofmeister et al. also composed the fivefold-twinned nanorods of silver [26]. Gryaznov *et al*. [27] reviewed the pentagonal symmetric particles and pentagonal needle-like crystals in nanostructured materials. Ding *et al*. [28] calculated and discussed the elastic deformation in a pentagonal rod with multiple twin boundaries. The pentagonal Cu NWs are formed by the extension of decahedron model particles.

Figures 2(a) shows pentagonal Cu NWs with diameters of 15 Å. Figure 2(b) shows the spreading sheets of the inside of pentagonal Cu NW shown in Fig. 2(a). Each sheet is composed of a square network in a {100} facet of fcc, excluding the central strand of atoms, and the wire axis always is along the <110> direction. Comparing with the fact that a shell in the CMS-type NWs is composed of a circular folding of the triangular network in a {111} facet of fcc, it is interesting that a shell in the pentagonal NWs is composed of a pentagonal folding of a square network. The pentagonal particles with multiple twin boundaries such as decahedral and icosahedral particles are composed of the tetrahedrons, which has only {111} surfaces. However, the subunits of the pentagonal NWs divided by five triangles show staking sequences oriented in the <100> directions. Theses looks like a quadrangular pyramids as shown in Fig. 2(c). Figure 2(c) shows the subunit indicated by a triangle Fig. 2(a). The distance of the nearest neighboring atoms is very slightly smaller than that of the fcc bulk, 2.556 Å. Kondo and Takayanagi [11] have shown that a straight, uniform nanobridge with the {111} planes along the <110> wire axis was stabilized and included both the square and the hexagonal lattices. Their results can help to understand the structural properties of the pentagonal NWs. They proposed a hexagonal prism model for the nanobridge, where the surface layers have a hcp lattices and the core has a fcc structure. Since pentagonal NWs are composed of a central atomic strand and pentagonal tubes made by fivefolding of {100} sheets and their subunits have a quadrangular-pyramid shape, the electron diffraction pattern of the pentagonal NWs shows both the square lattice and the hexagonal lattice [13]. For pentagonal NWs, the surface layers have a fcc structure and the core is a pentagonal structure. However, as the diameter of the pentagonal NW increases, a hexagonal lattice split from the core pentagon appears in the cross-section.

For CMS-type 16-11-6-1 Cu NW, the lattice constant in the direction of the wire axis is found to be 2.121 Å. The distances from the wire center to the atomic position in each shell are 6.494, 4.419, and 2.374 Å and the cohesive energies per atom of each shell are -3.030, -3.450, -3.465, and -3.386 eV, compared with -3.544 eV for the bulk material. For pentagonal 20-15-10-5-1 Cu NWs, the lattice constant in the direction of the wire axis is 2.462 Å. Distances from the wire center to atomic positions at the vertices of each pentagonal shell are 8.485, 6.345, 4.230, and 2.113 Å, and the average cohesive energies per atom of each shell are -2.994, -3.441, -3.485, -3.476, and -3.425 eV. In the present study, the average energies per atom (total cohesive energy per total atom number) for the CMS-type 16-11-6-1 and the pentagonal 20-15-10-5-1 NWs are -3.253 and -3.277 eV, respectively.

We also investigated the correlation functions of CMS-type and pentagonal Cu NWs to investigate the structural properties. Figure 3 shows the angular correlation function (ACF) and the pair correlation function (PCF) of pentagonal Cu NWs, compared to those of the CMS-type Cu

NW and fcc bulk at 300 K.   For pentagonal Cu NWs, the main peaks of the ACFs and the PCFs are in good agreement with those of the bulk. The ACFs and the PFCs of the CMS-type Cu NWs are different from those of the bulk in this work as well as in the previous work [4]. In the calculation of the ACFs, because we considered the nearest neighboring atoms, the small peaks related to the pentagonal structures appear at about 108º and 144º.

**Summary**

In summary, we simulated the tensile testing of cylindrical multi-shell Cu nanowire. Cylindrical multi-shell Cu nanowire was transformed into a pentagonal structure. Since the CMS-type NWs are not fcc structure, their structural properties are different from those of fcc structure [4]. However, although pentagonal Cu nanowires, too, are not a fcc structure, their structural properties are close to those of fcc structure, as the subunits have a quadrangular-pyramid shpae oriented in the <100> directions. Pentagonal NWs are composed of a central atomic strand and pentagonal tubes, five-times-folded {100} sheets, and a straight and uniform NWs with pentagonal planes along the <110> wire axis.

Figure Captions

Figure 1. Tensile force variations as a function of stretching. Structures when stretched to 7 Å are shown in bottom. The unit cell of pentagonal structure along the wire axis is shown in upper right side. The spheres of same color indicate atoms in the same layer.

Figure 2. (a) Pentagonal 20-15-10-5-1 NW. (b) The spreading sheets of the inside of Fig. 2(a). (c) The subunit indicated by a triangle in Fig. 2(a), such as a quadrangular pyramid oriented in the <100> direction.

Figure 3. Comparison with structural properties. The left and the right are the angular correlation and the pair distrubution function, respectively.

Figure

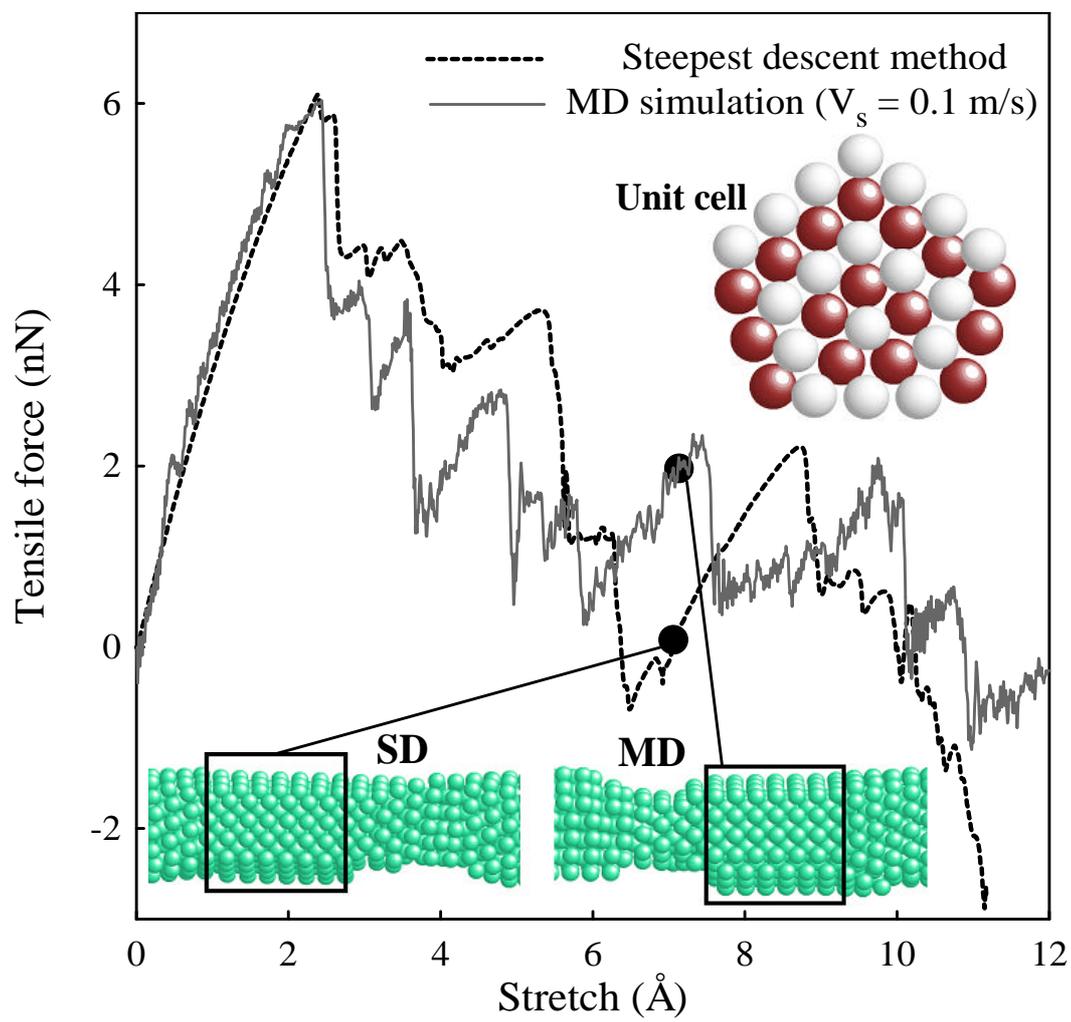

Figure 1.

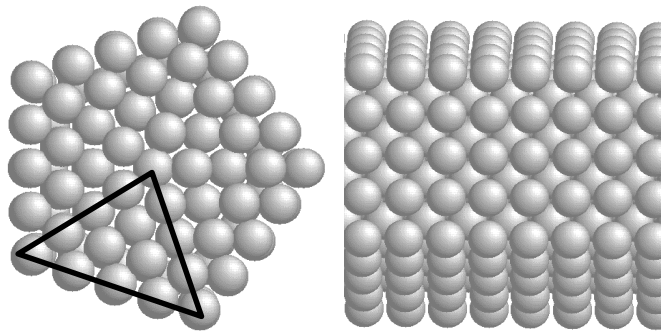
(a)

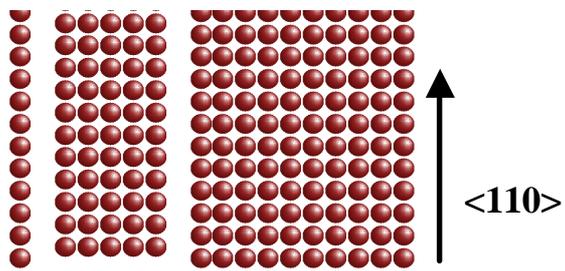
(b)

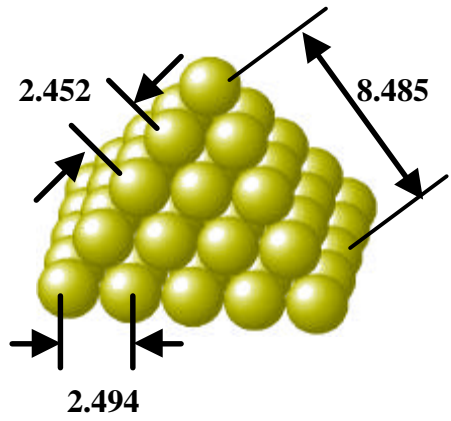
(c)

Figure 2.

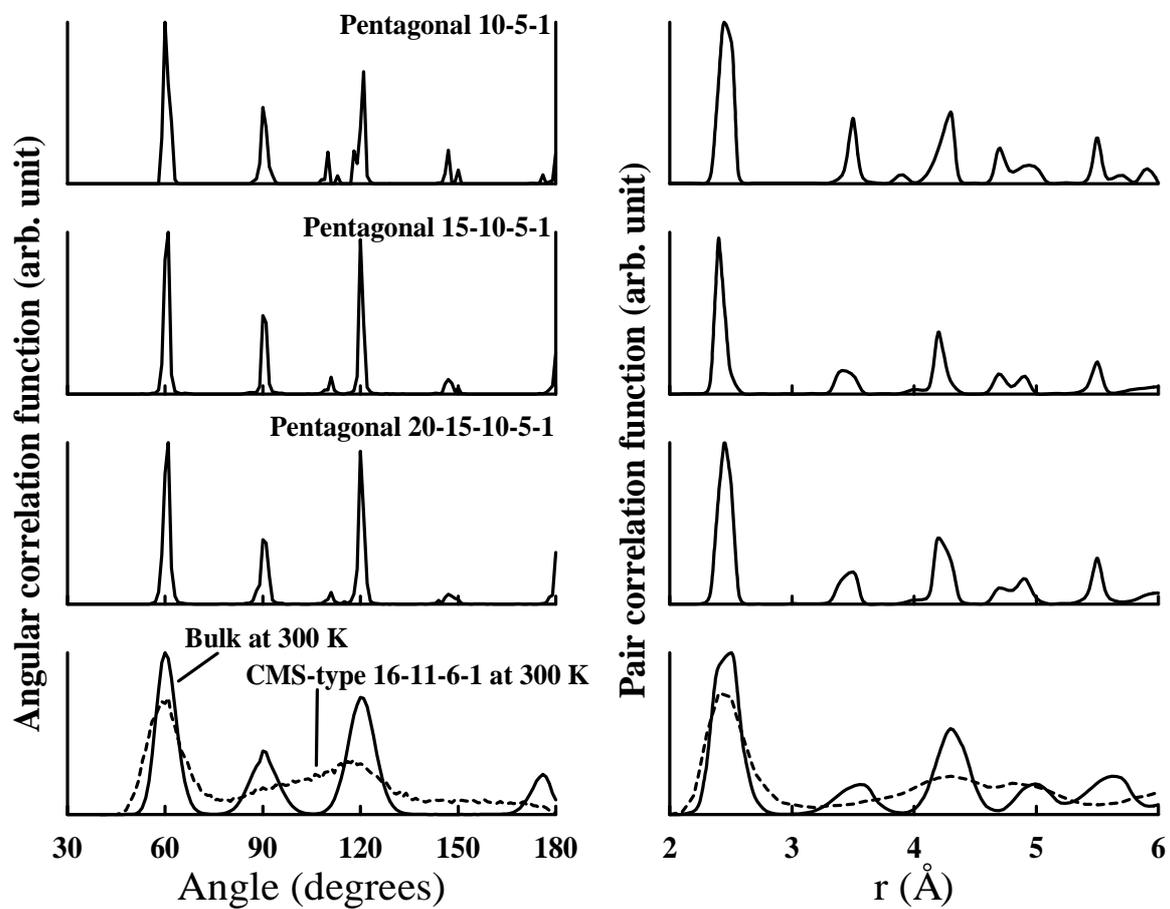

Figure 3.

15